\documentclass[journal]{IEEEtran}
\ifCLASSINFOpdf
\else
   \usepackage[dvips]{graphicx}
\fi
\usepackage[cmex10]{amsmath}

\usepackage{amssymb}
\usepackage{color}

\begin{document}
%
\title{De-Hashing: Server-Side Context-Aware Feature Reconstruction for Mobile Visual Search}
%
%
%

\author{Yin-Hsi Kuo
        and Winston H. Hsu

\thanks{Y.-H. Kuo is with the Graduate Institute of Networking and Multimedia, National Taiwan University (e-mail: kuonini@cmlab.csie.ntu.edu.tw). W. H. Hsu is with the Graduate Institute of Networking and Multimedia and the Department of Computer Science and Information Engineering, National Taiwan University, Taipei 10617, Taiwan (e-mail: whsu@ntu.edu.tw). Prof. Hsu is the contact person.}

}

%
%

\markboth{Journal of \LaTeX\ Class Files,~Vol.~13, No.~9, September~2014}%
{Shell \MakeLowercase{\textit{et al.}}: Bare Demo of IEEEtran.cls for Journals}
%



\maketitle

\begin{abstract}

Due to the prevalence of mobile devices, mobile search becomes a more convenient way than desktop search. Different from the traditional desktop search, mobile visual search needs more consideration for the limited resources on mobile devices (e.g., bandwidth, computing power, and memory consumption). The state-of-the-art approaches show that bag-of-words (BoW) model is robust for image and video retrieval; however, the large vocabulary tree might not be able to be loaded on the mobile device. We observe that recent works mainly focus on designing compact feature representations on mobile devices for bandwidth-limited network (e.g., 3G) \textcolor{black}{and directly adopt feature matching on remote servers (cloud).} However, the compact \textcolor{black}{(binary)} representation might fail to retrieve target objects (images, videos). Based on the hashed binary codes, we propose a de-hashing process that reconstructs BoW by leveraging the computing power of remote servers. To mitigate the information loss from binary codes, we further utilize contextual information (e.g., GPS) to reconstruct a context-aware BoW for better retrieval results. Experiment results show that the proposed method can achieve competitive retrieval accuracy as BoW while only transmitting few bits from mobile devices.

\end{abstract}


%
\IEEEpeerreviewmaketitle

\section{Introduction}
%
%
%
%
\IEEEPARstart{W}{ith} the explosive growth of mobile devices, the needs for mobile visual search are emerging. Because of the limited computing power and memory usage, it becomes a challenging problem for mobile visual search (MVS) \cite{Girod11}. Different from the traditional content-based image/video retrieval \cite{Sivic03}, \textcolor{black}{\cite{GuoPC15}}, mobile visual search requires lightweight computing and small data transmission. Hence, recent works focus on generating compact representations before transmitting the query. In order to achieve good retrieval accuracy, they will extract local features and compress them into binary codes for different applications, such as product search \cite{He12}, landmark retrieval \cite{Ji11}, \textcolor{black}{image/}video retrieval \textcolor{black}{\cite{LiuLZZT14a}, }\cite{Yeh14}, and interactive image exploring system \cite{Lu15}. Moreover, some works such as \cite{Chen14} further aim at on-device image matching.


 




\begin{figure}[!t]
\centering
\includegraphics[width=1\linewidth]{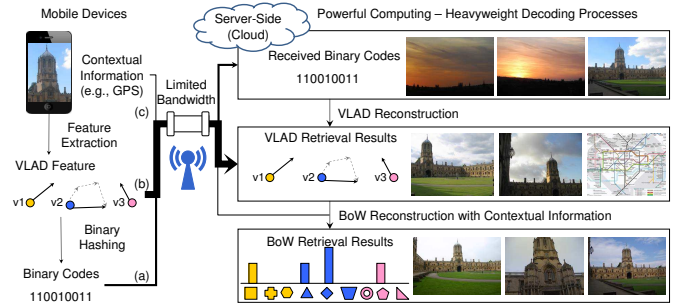}
\caption{For mobile visual search, the state-of-the-art approaches usually send hashed features through the bandwidth-limited network (3G) and apply feature matching on the server-side (cloud). We observe that we can utilize the computing power of remote servers to reconstruct a better feature representation from the hashed feature for mobile visual search. We propose a context-aware feature reconstruction to achieve better retrieval results. The thickness of the lines pass through the wireless network represents the amount of data transmission. We can transmit (a) binary codes or (b) a VLAD feature to the remote server. Meanwhile, we can integrate the proposed method with (c) contextual information to reconstruct a context-aware BoW.}
\label{fig:main_idea}
\end{figure}

The state-of-the-art visual feature---vector of locally aggregated descriptors (VLAD) \cite{Jegou12}---has been shown promising for image/video retrieval which has similar retrieval performance as bag-of-words (BoW) model \cite{Sivic03, delhumeau13}. \textcolor{black}{However, it might suffer from object queries \cite{arandjelovic13} because database images usually contain not only the target object but also cluttered backgrounds as shown in Figure~\ref{fig:main_idea}. These background features will affect the final representation of an image because VLAD aggregates features into a single representation (Section~\ref{subsec:VLAD_reconstruction}). For object retrieval, it is necessary to utilize BoW to mitigate the effect of noisy or cluttered backgrounds and provide better retrieval accuracy \cite{Chum11}. Otherwise, we need to generate and match multiple VLADs from different sizes of database images \cite{arandjelovic13}.} However, BoW requires a large vocabulary tree which \textcolor{black}{might not satisfy the memory constraint} on mobile devices. Hence, the authors in \cite{He12} propose bag of hash bits for image retrieval. They only consume a small amount of memory to generate compact binary codes for each local feature. Nevertheless, if the image/video contains many local features, the transmission time (cost) is still large.

\begin{table*}[!t]
\caption{The comparison of the state-of-the-art methods with our proposed method. Our proposed method only consumes a small amount of memory on mobile devices; moreover, \textcolor{black}{we can reconstruct context-aware feature representations from the received binary codes on remote servers and achieve better retrieval accuracy.}} \label{table:related_work}
\vspace{-10pt}
\fontsize{7pt}{8pt}\selectfont
\begin{center}
\begin{tabular}[c]{l||c|c|c|c}
\hline%
        & Transmission (KBytes) & Transmission Size & Memory Consumption on Mobile (KBytes) & Meta-Information \\ \hline \hline %
BoW (1M) \cite{Chen09}                & 0.36 - 0.6 & Dependent ($\sim$13 bits/feature) & 68,000 (quantization tree) & 8-bit per dimension \\
CHoG \cite{Chandrasekhar12}  & 0.38 - 5.3 & Dependent ($\sim$60 bits/feature) & - & Feature location  \\ %
Product Search \cite{He12} & 0.5 - 5  & Dependent (80 bits/feature) & 20 (64x80x4, projection matrix) & Boundary information \\
REVV \cite{Chen11b, Chen2013}  & - (0.35) & Independent & 264 (tree + matrix) & On-device matching \\ 
MCVD \cite{Ji11}                  & 0.025 - 0.4 & Independent & 690 - 3,440 & Contextual information \\ %
IMShare \cite{Dai12}                  & 8 (2 + 6) & Dependent & - & Thumbnail \\ \hline %
Our proposed method                   & 0.8 - 1.6 & Independent & 45 - 121.9 (tree + matrix) & Contextual information \\ \hline
\end{tabular}
\end{center}
\vspace{-10pt}
\end{table*}






Compared to mobile devices, remote servers (cloud) have stronger
computing power and storage. We find that the state-of-the-art approaches usually apply feature matching directly on the received features. To leverage the computing power of remote servers, we utilize VLAD as an intermediate feature and generate a better feature representation before feature matching. To tackle this challenge, we observe the relation between VLAD and BoW (Section \ref{sec:VLAD2BoW}), and propose to estimate possible visual words (VWs) from VLAD on remote servers. Meanwhile, motivated by the on-device image matching \cite{Chen14}, we generate compact and fixed length binary codes from VLAD on mobile devices. As Figure \ref{fig:main_idea} shows, given a query (image or video), we adopt a lightweight encoding process that hashes features into binary codes on mobile devices, and design a novel decoding process on the server-side called \textit{de-hashing}. \textcolor{black}{To the best of our knowledge, this is the first work that attempts to reconstruct BoW from VLAD or binary codes.} Hence, the proposed method only transmits a single compact representation (with contextual information) from a huge amount of local features in the mobile query.

\textcolor{black}{In this work, we aim to leverage both the mobile- and cloud-based configurations (two heterogeneous platforms) for balancing effectiveness and efficiency for image/video retrieval (also extendable for image/video classification). In a novel way, we exploit the (unbalanced) computation capabilities between the two heterogeneous platforms (i.e., mobile devices and cloud servers) and seek new feature learning and representations friendly with the whole path from mobiles, through communication channel, and to the cloud.} The primary contributions of the paper include,

\begin{itemize}

\item Proposing a de-hashing process that leverages contextual information to approximate BoW from a compact feature representation on the server-side (Section \ref{sec:VLAD2BoW}).

\item Investigating the memory consumption of binary hashing on mobile devices (Section \ref{sec:binary2VLAD}).

\item Conducting experiments on two retrieval benchmarks and demonstrating that the proposed method can achieve better retrieval accuracy compared to the original binary codes (Section \ref{sec:experiments}).

\end{itemize}

\section{Related works}

We aim to provide better search results for mobile visual search; hence, we introduce recent works and compare them with our proposed method. The state-of-the-art image/video retrieval systems usually extract BoW from a query; however, it might be infeasible to transmit the query image or video from mobile devices under the unstable network connection \cite{Girod11}. Recent works focus on low bitrate mobile visual search \cite{He12, Chen14}\textcolor{black}{, \cite{Chen15}}. They propose a lightweight encoding process on the extracted features before transmission. These approaches can roughly divide into four compression methods---BoW-based \cite{Chen09}, CHoG-based (compressed histogram of gradients) \cite{Chandrasekhar12}, hash-based \cite{He12}, and VLAD-based \cite{Chen2013} methods.

For BoW-based method, they attempt to prune the large vocabulary tree so that it can fit in the mobile memory \cite{Ji11} and apply standard encoding methods (e.g., run-length encoding or arithmetic coding) to further compress the BoW histogram \cite{Chen09}. Instead of applying BoW, the authors in \cite{Chandrasekhar12} propose a novel descriptor---CHoG---by considering the limited resources on mobile devices. Different from vector quantization (or tree-based) approach, hash-based method utilize a small amount of memory consumption (projection matrix) to efficiently generate compact binary codes \cite{He12, Zhou14}\textcolor{black}{, \cite{LiuLZZT14, Wang16}}. Moreover, recent works further propose more compact and discriminative binary descriptors \cite{Trzcinski13a}\textcolor{black}{, \cite{Trzcinski15}} to achieve similar performance as floating-point descriptors. These approaches have shown promising results on different benchmarks; however, the transmission cost highly depends on the number of extracted local features. For high-resolution images, it might exceed 1,000, but these methods usually extract few hundreds of features to perform image or video matching.

To tackle this problem, the authors in \cite{Chen2013} propose a novel compact global signature called residual enhanced visual vector (REVV) which compresses VLAD feature into binary codes. Hence, their proposed method only needs few bits to represent each image. It is very suitable for on-device image retrieval for personal photos because the whole process can be done on the mobile device. Similarly, the authors in \cite{Ji11} propose a multiple-channel coding based compact visual descriptor (MCVD), which compresses the BoW histogram into a reversible binary signature on mobile devices, and restore MCVD to BoW in a lossy manner on the remote server. However, their method needs to retain different compression functions for different locations on mobile devices. Instead of utilizing individual compression function, we attempt to provide a universal compression method on mobile devices and investigate different reconstruction methods with contextual information adaptively on remote servers.

Rather than focusing on mobile devices, recent works utilize the computing power of remote servers (cloud) to reconstruct an image from its (compressed) local features \cite{Dai12, Weinzaepfel11, Yue13}, or generate distinctive image representations \cite{Liu16} \textcolor{black}{, \cite{Hong15}}. Motivated by aforementioned works, our proposed framework considers both the limited resources on mobile devices and the stronger computing power on remove servers for mobile visual search. Table \ref{table:related_work} shows the overall comparison of the proposed method and prior works. We only consume a small amount of memory to generate binary codes, and further reconstruct them into a context-aware BoW for better retrieval results.

\begin{figure*}[!t]
\centering
\includegraphics[width=1\linewidth]{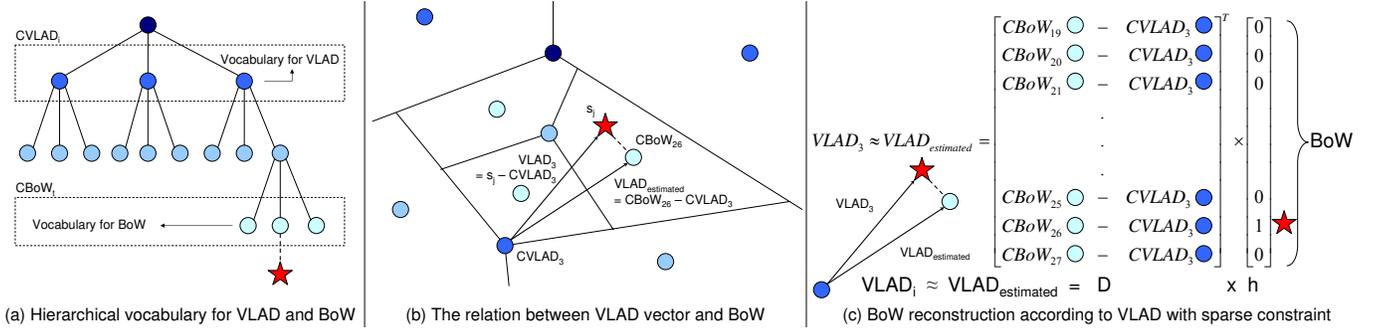}
\caption{Illustration of BoW reconstruction from VLAD. (a) We utilize a hierarchical vocabulary tree to maintain both VLAD and BoW centers. Therefore, given a local feature (rad star), we can estimate its possible BoW center (light blue circle) from its VLAD center (dark blue circle). (b) Due to the large vocabulary of BoW, \textcolor{black}{we propose to calculate $VLAD_i$ by replacing the local feature (red star) with a BoW center}. (c) Hence, with the connection between BoW and VLAD, we can roughly approximate BoW from VLAD.} \label{fig:vlad2bow}
\end{figure*}

\section{Context-Aware BoW Reconstruction} \label{sec:VLAD2BoW}

\textcolor{black}{To leverage both the mobile- and cloud-based configurations for mobile visual search}, we utilize VLAD as an intermediate feature for compression and reconstruction. This is because VLAD only requires a small amount of memory for quantization tree which is much smaller than BoW model (the fourth column in Table \ref{table:related_work}) and is suitable for mobile devices.\footnote{In our prototype, it takes 7.6 milliseconds to extract VLAD from SURF \cite{Bay08} for a 320 x 240 image in iPhone 5.} Hence, we aim to reconstruct BoW from VLAD for better retrieval accuracy (e.g., object queries). By observing the relation between VLAD and BoW in the hierarchical vocabulary tree, we are able to approximate BoW from VLAD detailed in Section \ref{subsec:VLAD_reconstruction}. Besides, we can utilize contextual information to reconstruct different BoW histograms from the same VLAD, or obtain a better BoW even if the VLAD is approximated from binary codes (cf. Section \ref{sec:binary2VLAD}). We investigate different reconstruction approaches in Section \ref{subsec:dictionary_selection}. Moreover, we integrate the reconstruction method with a prior knowledge from the initial search results for better BoW reconstruction in Section \ref{subsec:reconstruction_prior}.

\subsection{BoW Reconstruction from VLAD with Sparse Constraint} \label{subsec:VLAD_reconstruction}

To reconstruct BoW from VLAD, we observe that they can be connected by their vocabulary trees. As shown in Figure \ref{fig:vlad2bow}(a), we utilize a hierarchical vocabulary (e.g., HKM, hierarchical k-means \cite{Nister06}) to construct the relation between VLAD and BoW. The top-layered vocabulary is used for VLAD and the leaf nodes are BoW vocabulary. Therefore, if a local feature (red star) belongs to a VLAD center, we are able to estimate possible BoW centers (VWs) it belongs to (i.e., the sub-tree of the VLAD center).

We first define the generation process for VLAD.  For each $VLAD_i$, we collect all the local features ($s_j$) quantized into the same VLAD center ($CVLAD_i$), and aggregate the difference between local features and the center as
\begin{align} \label{eq:VLAD}
VLAD_i = \sum_{s_j \in CVLAD_i} (s_j-CVLAD_i). 
\end{align}
 As an example shown in Figure \ref{fig:vlad2bow}(b), we assume the original local feature is the red star and circles are codeword centers. Different colors represent different levels of hierarchical vocabulary. Hence, for each local feature ($s_j$), it will contribute to $VLAD_i$ by the difference between the red star and the dark blue circle ($s_j-CVLAD_i$). The final VLAD will concatenate all the $VLAD_i$ into a single feature.

\textcolor{black}{By observing large vocabulary of BoW for image retrieval, we propose to substitute a BoW center ($CBoW_t$, light blue circle near the red star) for the local feature (red star).} In other words, the difference ($s_j-CVLAD_i$) of $VLAD_i$ can be approximated by the light blue circle minus the dark blue circle ($CBoW_t-CVLAD_i$). Therefore, as shown in Figure \ref{fig:vlad2bow}(c), $VLAD_i$ (Eq. \eqref{eq:VLAD}) can be approximated by $Dh$, where $h$ represents BoW histogram and $D$ is generated by the difference between BoW centers (sub-tree of $CVLAD_i$) and VLAD centers (i.e., $D=[CBoW_t-CVLAD_i]^T$).

Meanwhile, BoW histogram is usually a sparse vector because the number of local features is relatively small. The sparse constraint also provides an opportunity to correctly reconstruct BoW histogram from VLAD. Therefore, given $VLAD_i$ ($v$), BoW ($h$) reconstruction can be formulated as
\begin{align} \label{eq:formulation_BoW_sparsity}
f_{h} = \min_h \|v-Dh\|_2^2 + \lambda \|h\|_{1}, \text{ subject to } h > 0, 
\end{align}
where the first term measures the reconstruction quality between $VLAD_i$ ($v$) and approximated VLAD ($Dh$). $\lambda$ is a regularization parameter that controls the sparsity of BoW ($h$). The formulation is similar to the sparse coding problem \cite{Lee06} but the dictionary ($D$) is pre-defined according to the difference between the centers of BoW and VLAD. Hence, we do not need to train the dictionary and can utilize SPArse Modeling Software (SPAMS) \cite{Mairal09, Mairal10} for solving the above formulation. Note that this idea is also similar to compressive sensing that reconstructs sparse signals \cite{Baraniuk07}.

\subsection{Context-Aware Dictionary Selection (CADS)}
\label{subsec:dictionary_selection}

\begin{figure}[!t]
\centering
\includegraphics[width=1\linewidth]{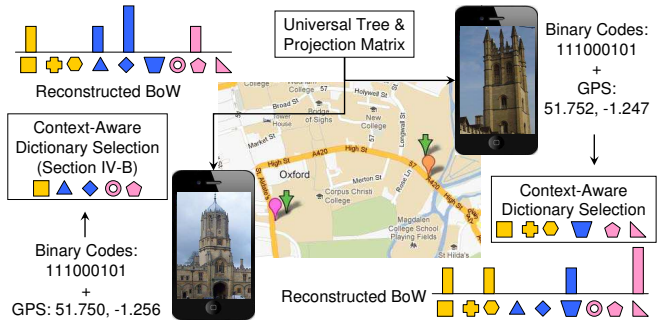}
\caption{Illustration of dictionary selection based on different contextual information. We utilize the same projection matrix and small quantization tree on mobile devices, and reconstruct different BoW representations on the server-side according to the contextual information (e.g., GPS).} \label{fig:DGPS}
\end{figure}

It is reasonable since mobile is augmented with geo-information; hence, we further propose a context-aware dictionary selection (CADS) for BoW reconstruction. By utilizing contextual information, we are able to reconstruct different BoW histograms from the same VLAD or binary codes as shown in Figure \ref{fig:DGPS}. We utilize a single and universal vocabulary tree for BoW reconstruction; however, \textcolor{black}{we dynamically select possible dictionary ($D_{context}$ in Eq.~\eqref{eq:formulation_BoW_sparsity}) based on different contextual cues. Instead of using all VWs (e.g., 9 in Figure~\ref{fig:vlad2bow}), we only retain few relevant VWs (e.g., top-ranked candidates by binary codes or GPS similarity). Hence, we consider both visual and contextual information for generating more discriminative BoW representations. Note that we can record possible VWs (sub-vocabularies) in an offline manner, or select them on the fly (e.g., calculating GPS similarities). Moreover, the selection process also leads to faster solving time because the hypotheses of VWs are greatly reduced.}

\subsection{BoW Reconstruction with Prior Knowledge (BRPK)} \label{subsec:reconstruction_prior}

In addition to context-aware dictionary selection, we further utilize initial ranking results as prior knowledge to estimate a more powerful BoW representation. We know that the re-ranking process is a common query expansion method to provide a better performance. Hence, \textcolor{black}{we generate a pseudo-BoW} from top-ranked search results (i.e., obtained from binary codes in our experiments).\footnote{\textcolor{black}{By considering the limited bandwidth for MVS, we only transmit a small amount of data (binary codes) to remote servers. It is efficient to retrieve images via binary codes. Nevertheless, we can also transmit VLAD and retrieve superior ranking results for better BoW reconstruction.}} Based on the prior knowledge, our goal is to reconstruct BoW representation which is similar to the approximated VLAD \textcolor{black}{(Section~\ref{sec:VLAD_R})} and the pseudo-BoW. Hence, the reconstruction operation can be formulated as
\begin{align} \label{eq:formulation_prior}
f_h = \min_h \alpha \frac{\|v-Dh\|_2^2}{N_{1}} + (1-\alpha) \frac{\|h-h_{0}\|_2^2}{N_{2}}, 
\end{align}
where $\alpha$ stands for the influence between the first and the second terms. $N_{1}=\|v\|_2^2$ and $N_{2}=\|h_0\|_2^2$ are the normalization terms. The first term is followed by the previous section to reconstruct BoW representation. The second term is to ensure the reconstructed BoW is similar to the pseudo-BoW ($h_0$) that is generated from top-ranked results. By considering two different criteria, the proposed BoW reconstruction with prior knowledge (BRPK) can obtain a better BoW for MVS.\footnote{Note that here we choose L2 regularization is to speed up the reconstruction process because the response time for image retrieval systems is also an important factor. Hence, with prior knowledge from ranking results, we can roughly estimate possible VWs and round down the reconstructed BoW histogram to enforce the sparsity in the final BoW representation.}


To solve the proposed formulation, we find that it is equivalent to the generalized Tikhonov regularization and exists an optimal solution. Hence, we are able to compute the unique solution $h^*$ of Eq. \eqref{eq:formulation_prior} analytically. Let $\alpha_1 = \frac{\alpha}{N_{1}}$ and $\alpha_2 = \frac{1-\alpha}{N_{2}}$, a direct solution would lead to
\begin{align}
h^* = (\alpha_1 D^TD + \alpha_2 I)^{-1}(\alpha_1 D^Tv + \alpha_2 h_0). %
\label{eq:soluation_h}
\end{align}
However, if the inverse matrix is large (e.g., $10,000 \times 10,000$, possible VWs), it is time-consuming to compute the solution. We can transform $D^{T}D$ to $DD^{T}$ which is related to the feature dimension (e.g., $128 \times 128$, SIFT \cite{Lowe04}). The transformation \cite{Petersen12} is based on the identity of the inverse function $(I+AB)^{-1}A=A(I+BA)^{-1}$ and $(I+AB)^{-1}=I-A(I+BA)^{-1}B$. Then, we can re-write Eq. \eqref{eq:soluation_h} as
\begin{eqnarray}
h^* & = & \alpha_1 D^T(\alpha_1 DD^T + \alpha_2 I)^{-1}v \nonumber \\
& + & [I - \alpha_1 D^T(\alpha_1 DD^T + \alpha_2 I)^{-1}D]h_0 \nonumber \\
& = & \alpha_1 D^T(\alpha_1 DD^T + \alpha_2 I)^{-1}(v - Dh_0) + h_0.
\end{eqnarray}
We can efficiently reconstruct BoW histogram based on the above solution. Moreover, \textcolor{black}{we can achieve better retrieval accuracy} because we not only consider the initial ranking results but also utilize both visual and contextual information.

\begin{figure*}[!t]
\centering
\includegraphics[width=1\linewidth]{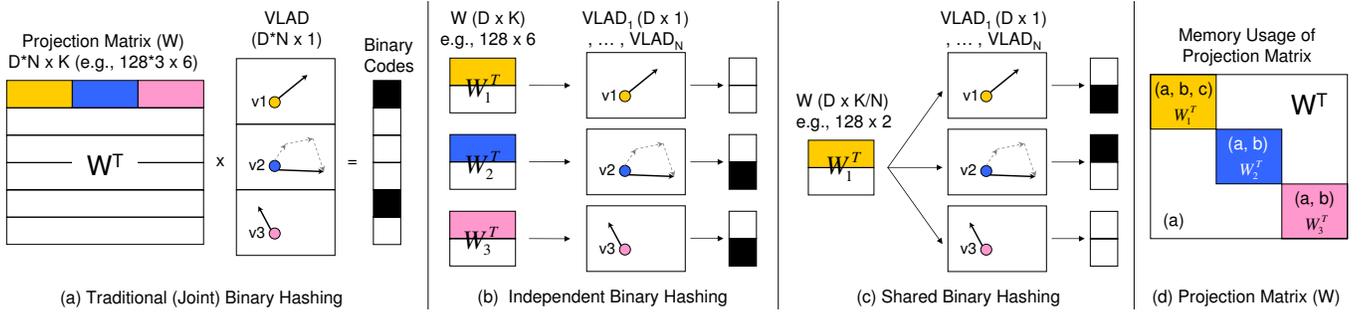}
\caption{Illustration and comparison of different binary hashing methods. (a) The traditional (joint) binary hashing \textcolor{black}{projects the original (high-dimensional) feature space onto a reduced feature space. However, memory usage is proportional to the original and reduced features' dimensionality.} (b) We can split the original feature (VLAD) into sub-features ($VLAD_i$) so that the projection matrix can be reduced. (c) Moreover, we can learn a uniform (shared) projection matrix for all the sub-features. The shared method needs more bits to achieve competitive retrieval accuracy compared to the traditional one (Section~\ref{sec:exp_bow}); hence, it is a trade-off between the performance and memory usage. (d) The memory usage of projection matrix on three approaches. The symbols in different rectangles represent the consumed memory. For example, the shared method (c) only requires the top-left (yellow) rectangle for the projection matrix.} \label{fig:binary_hashing}
\end{figure*}

\section{Reversible Binary Code Generation} \label{sec:binary2VLAD}

Under the unstable wireless network, transmitting VLAD (e.g., 8,192 dimensions, 32KB) back to the remote server might still be infeasible. To further compress VLAD, we apply hash-based methods to generate binary codes on mobile devices for reducing the transmission cost. The binary hashing functions ($bh_k(.)$) can be formulated as
\begin{align} \label{eq:hashing}
bh_k(x) = (sgn(w_k^Tx)+1)/2, k = 1, ... , K,
\end{align}
where $x \in \mathbb{R}^d$ (zero mean) is the original (visual) feature and $w_k \in \mathbb{R}^d$ is a \textcolor{black}{projection vector} sampled from Gaussian distribution (\textit{random projection, RP}) \cite{Charikar02} or learned from training data \cite{Wang12}. $K$ is the total number of hashing functions (to generate $K$ bits for each feature). \textcolor{black}{Although it is compact for transmission, the limited memory of mobile devices enforces the projection matrix ($W \in \mathbb{R}^{d \times K}$) should be small as well.} Hence, we introduce principal component analysis (PCA) hashing in Section \ref{sec:PCA_hashing} and investigate various ways for generating compact codes on mobile devices in Section \ref{sec:memory_efficient}. In Section \ref{sec:VLAD_R}, \textcolor{black}{we reverse binary codes for obtaining approximated VLAD which is used for BoW reconstruction.} 

\subsection{Principal Component Analysis Hashing (Joint PCAH)} \label{sec:PCA_hashing}

Recent works demonstrate that PCA hashing (PCAH) provides high binarization quality and retrieval performance \cite{Gong12, Jegou12PCA}. The projection matrix ($W$) of PCAH is learned from the covariance matrix ($XX^T$) of training data ($X$). By selecting the largest $K$ eigenvectors to form $W$, we are able to generate binary codes based on Eq.~\eqref{eq:hashing}.\footnote{\textcolor{black}{We can apply other binarization or (semi-) supervised hashing \cite{Wang12} strategies, or compute weights for each bit (dimension) to improve retrieval accuracy \cite{Jiang13}.} However, we are to investigate BoW reconstruction from binary codes so we only utilize the standard PCAH in our experiments.} However, the projection matrix might be very huge if we directly apply PCAH on the high-dimensional feature space \textcolor{black}{called \textit{joint PCAH} in our experiments. For example, to generate 1,024-bit binary codes ($K$) from 12,800-d VLAD ($D*N$), the projection matrix requires around 50MB (i.e., 12,800 x 1,024 x 4 bytes, $D*N \times K$ in Figure \ref{fig:binary_hashing}(a)) for memory usage.}\footnote{$D$ is the dimension of local features (e.g., SIFT, SURF), and $N$ is the number of VLAD centers. $D*N$ means the concatenated dimension of VLAD.} As a result, the memory consumption might be similar to 1M vocabulary tree (e.g., 128MB = 1M centers x 128 bytes). In other words, \textcolor{black}{if we increase the dimension of the original or reduced features for better retrieval accuracy,} the memory cost will be infeasible on mobile devices. Hence, we further consider memory-efficient binary hashing for mobile visual search.

\subsection{Memory-Efficient Binary Hashing (Ind. and Shared PCAH)} \label{sec:memory_efficient}

Followed by product quantization \cite{Jegou11PQ} and on-device image matching \cite{Chen2013}, \textcolor{black}{we split high-dimensional feature (VLAD) into sub-features ($VLAD_i$) and learn the projection matrix individually called \textit{independent (ind.) PCAH}}. Moreover, we can learn a shared projection matrix for all the $VLAD_i$ \textcolor{black}{called \textit{shared PCAH}. As shown in Figure \ref{fig:binary_hashing}(b) and \ref{fig:binary_hashing}(c), the memory cost of ind. PCAH and shared PCAH is $D \times K$ and $D \times K / N$, respectively.} Figure \ref{fig:binary_hashing}(d) shows the overall comparison of memory consumption by different approaches. \textcolor{black}{Compared to others, the shared approach only requires relatively small memory usage ($W_1^T$, top-left yellow rectangle).} It is a trade-off between the performance (bit length) and memory consumption on mobile devices. Note that these approaches are complementary to sparse learning (e.g., sparse PCA \cite{Journee10}), \textcolor{black}{bilinear \cite{Gong13}, or circulant approaches \cite{Yu14}. We can combine them with our proposed method} to further reduce the memory consumption. Besides, as mentioned in \cite{Perronnin10}, a simple component-wise sign binarization on VLAD can achieve good results. Hence, we will also discuss the results in Section~\ref{sec:exp_bow}.

\subsection{VLAD Approximation from Binary Codes} \label{sec:VLAD_R}

\textcolor{black}{Based on the previous section, we can efficiently generate binary codes via PCA-based methods (i.e., $W^Tx$) on mobile devices. To reconstruct BoW from VLAD on remote servers (Section~\ref{sec:VLAD2BoW}), we have to reverse the retrieved binary codes and obtain approximated VLAD for the reconstruction. PCA-based methods have the orthogonal projection matrix property ($W^TW=I$); therefore, we can reverse binary codes by multiplying the same projection matrix (i.e., $x \sim WW^Tx$).} However, for binary hashing (i.e., $sgn(W^Tx)$), we will lose more information due to the binarization process. This is also the reason why we propose to utilize contextual information for better BoW reconstruction (Section \ref{subsec:dictionary_selection}). Besides, as mentioned in \cite{Jegou12, Gong12}, we can apply an orthogonal transformation (e.g., RR, random rotation) to mitigate the \textcolor{black}{quantization error of binarization}, and obtain better binary codes and the reversed feature. It is essential for the high-dimensional feature; hence, we will apply random orthogonal rotation on joint PCAH (\textit{joint PCAH-RR}) in Section~\ref{sec:exp_bow}.

\section{Experiment Results and Discussions} \label{sec:experiments}

\subsection{Experiment Setting}

In our experiments, we construct 1M hierarchical vocabulary tree with 6 levels and 10 branches \cite{Nister06} for BoW model, and choose the second level for VLAD (i.e., 100 centers). The distance measurement is L1 for BoW and L2 for VLAD in the retrieval process. We conduct our proposed method on two datasets. For mobile scenario, we choose \textbf{Stanford mobile visual search (SMVS) dataset} \cite{Chandrasekhar11, SMVSD} which contains 1,193 single reference images with 3,269 mobile queries \textcolor{black}{across 8 image categories (i.e., CD, DVD, Books, Video Clips, Landmarks, Business Cards, Text documents, Paintings).} We resize images to small resolution (maximum height or width is 640) and use speeded-up robust features (SURF) \cite{Bay08} on mobile devices and generate VLAD with 6,400 dimensions. The evaluation metric is recall at $Num$ (R@Num) and NDCG. \textcolor{black}{We set the relevance score of ground truth images to be 1. Each query only has one reference (correct) image; hence, the ideal DCG is 1 and NDCG is equivalent to $1/log_2(r+1)$, where $r$ is the rank number of the reference image in the retrieval results.} As reported in \cite{Chen2013}, BoW and REVV can achieve around 75\% recall with spatial verification on top 50 candidates. To demonstrate the effect of our proposed method, we do not apply spatial verification on the image search results (i.e., evaluating on the initial ranking results).

Moreover, for demonstrating the effect of image object retrieval, we choose \textbf{Oxford buildings (Oxford) dataset} \cite{Philbin07} which contains 5,062 images with 55 object queries. We use rootSIFT \cite{Lowe04, Arandjelovic12} to generate 12,800-d VLAD with intra-normalization~\cite{arandjelovic13}.\footnote{Note that the proposed method can also apply to video retrieval because we can extract features from each frame and aggregate them to form VLAD.} In order to provide contextual information on Oxford, we generate GPS for each image followed by \cite{Chen11} that utilizes Gaussian error model to approximate GPS from the true location. As mentioned in \cite{Philbin07}, the dataset is downloaded from 17 Flickr queries (keywords). Hence, we can obtain 16 buildings' GPS information (true location) from Wikipedia. The remaining ``Oxford'' keyword is randomly assigned to the other 16 keywords. The evaluation metric is mean average precision (MAP). The MAP of VLAD on Oxford is 0.371, and is similar to \cite{Jegou12} which reports the MAP of VLAD with 4,096 dimensions is 0.304. The MAP of BoW and GPS on Oxford is 0.483 and 0.168, respectively. As reported in \cite{Philbin07}, the MAP of SIFT with hierarchical vocabulary tree (HKM) for BoW is 0.439. In our experiments, the original SIFT can achieve 0.422 which is similar to their paper. Hence, we utilize the more robust rootSIFT (0.483) for constructing vocabulary in our experiments. \textcolor{black}{Meanwhile, for evaluating the effect of training data, we further utilize two additional datasets---\textbf{Paris \cite{Philbin08}} and \textbf{Landmarks \cite{Babenko14}}---for learning the vocabulary tree and projection matrix. For Landmarks dataset, we randomly sample 30 images from 721 landmarks, and utilize around 22,000 images for training.}

\begin{figure}[!t]
\centering
\includegraphics[width=0.7\linewidth]{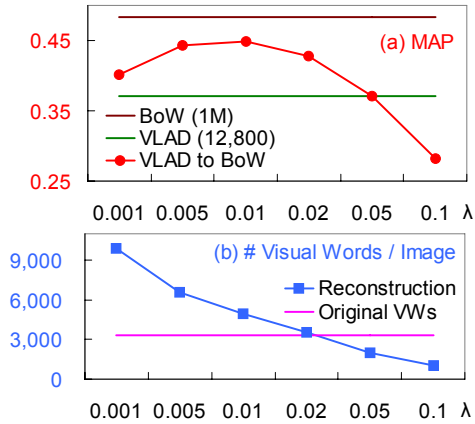}
\caption{Parameter sensitivity of $\lambda$ in Eq. \eqref{eq:formulation_BoW_sparsity} on Oxford buildings dataset. \textcolor{black}{(a)} It shows that we can achieve similar results as the original BoW when selecting a proper value of $\lambda$. \textcolor{black}{(b)} Moreover, the value should be adjusted by the number of VWs we are to reconstruct. For Oxford dataset, the averaged number of VWs is around 3,350. Hence, we set $\lambda = 0.02$ in the experiments.} \label{exp:vlad2bow}
\end{figure}

\subsection{Experiments on the Choice of Lambda ($\lambda$)}

The most important parameter of our proposed context-aware BoW reconstruction is $\lambda$ in Eq. \eqref{eq:formulation_BoW_sparsity} because it will affect the reconstruction quality for image retrieval. Hence, we conduct experiments with different values of $\lambda$ on Oxford dataset. As Figure \ref{exp:vlad2bow}\textcolor{black}{(a)} shows, the reconstructed BoW (VLAD to BoW) has similar retrieval accuracy as BoW when $\lambda$ ranges between 0.005 to 0.02. We find that the number of reconstructed VWs (6,500 to 3,500) is larger than the original VWs (around 3,350) \textcolor{black}{as shown in Figure \ref{exp:vlad2bow}(b).} \textcolor{black}{The reconstruction step may include noisy VWs; however, it can also be viewed as a by-product of soft (multiple) assignment \cite{Philbin08}.} When $\lambda$ is small (0.001), we reconstruct too many noisy VWs and the retrieval accuracy decreases. Conversely, if $\lambda$ is larger than 0.05, the MAP is worse than VLAD because we only reconstruct few VWs and it is hard to retrieve similar images.

\textcolor{black}{We observe that a proper value for $\lambda$ is related to image resolution because the number of extracted local features (VWs) depends on it.} Therefore, for Oxford dataset with 1,024 x 768 pixels, we set $\lambda$ to be 0.02 \textcolor{black}{because the averaged number of VWs is around 3,350. Note that the choice of $\lambda$ for Oxford is not based on the highest MAP ($\lambda=0.01$). Similarly,} based on the experiments in Figure \ref{exp:vlad2bow}(b), for SMVS dataset with small resolution (maximum height or width is 640), we can directly adjust the value of $\lambda$ to 0.1 without evaluating the performance (i.e., viewing Oxford as an independent dataset for SMVS) because the averaged number of VWs is around 830. In other words, we can decide $\lambda$ based on the desired number of reconstructed VWs as shown in Figure~\ref{exp:vlad2bow}(b).

\begin{figure}[!t]
\centering
\includegraphics[width=1\linewidth]{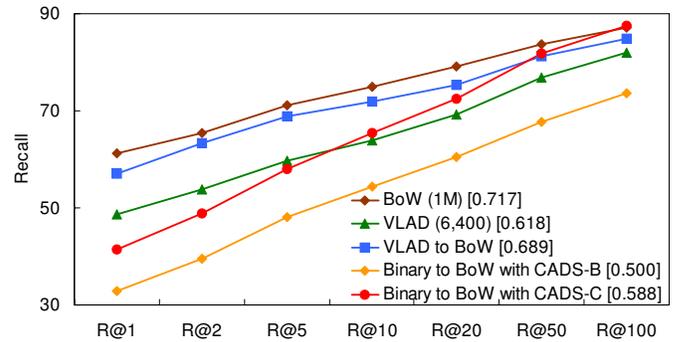}
\caption{BoW reconstruction from the original retrieved feature (i.e., VLAD or binary codes) on Stanford mobile visual search dataset. It shows that the reconstructed BoW (VLAD to BoW) can achieve similar retrieval accuracy as BoW. However, if we apply context-aware dictionary selection based on binary ranking results to estimate BoW from binary codes (Binary to BoW with CADS-B), the retrieval accuracy is worse than VLAD and BoW because the binarization step loses too much information. By utilizing additional contextual information (category types) to select the most possible VWs as mentioned in Section \ref{subsec:dictionary_selection}, we can achieve better retrieval results. \textcolor{black}{The values in the square brackets represent NDCG scores.}} \label{exp:SMVS_v2b_b2b}
\end{figure}

\subsection{BoW Reconstruction on SMVS Dataset} \label{sec:exp_bow}

First, we conduct experiments on SMVS dataset. As Figure~\ref{exp:SMVS_v2b_b2b} shows, the retrieval accuracy of BoW (even the reconstructed BoW) is better than VLAD \textcolor{black}{(0.717 or 0.689 vs. 0.618)}. The results confirm that it is necessary to reconstruct BoW from VLAD on the server-side. As mentioned in Section \ref{sec:VLAD2BoW}, we can reconstruct BoW from VLAD or (reversible) binary codes. The blue curve (rectangle) in Figure \ref{exp:SMVS_v2b_b2b} shows the reconstructed BoW (VLAD to BoW) can achieve competitive retrieval accuracy as the original BoW. This represents that the proposed method can successfully approximate the original BoW. However, if we reconstruct BoW directly from binary codes and utilize the approximated VLAD for the reconstruction, the results might be worse than the original binary codes because the binarization process loses too much detailed information to reconstruct the original VLAD.\footnote{For better reconstruction results from binary codes to VLAD, we also adopt iterative quantization (ITQ) as proposed in \cite{Gong12}.}

\textcolor{black}{In order to demonstrate the effect of contextual information for BoW reconstruction, we assume the class information is known (i.e., book, cd, painting, etc., given by SMVS dataset) for context-aware dictionary selection. For real applications, we can apply classification for obtaining possible class information. Hence, for fair comparison, we only utilize GPS information and binary codes for Oxford dataset in Section~\ref{sec:exp_bow} and Table~\ref{table:numberOfBits}. As shown in Figure \ref{exp:SMVS_v2b_b2b}, when utilizing binary ranking results as a contextual cue for BoW reconstruction (Binary to BoW with CADS-B), the retrieval accuracy is worse than the original VLAD. However, when utilizing class information as an additional cue (Binary to BoW with CADS-C), we can achieve better retrieval accuracy (0.5 to 0.588). This is because the selection process can filter out those impossible (irrelevant) VWs. Note that we only utilize contextual information for BoW reconstruction (i.e., selecting $D_{context}$ in Section \ref{subsec:dictionary_selection}); hence, the proposed method can further combine with other re-ranking methods (e.g., late fusion) in the retrieval process.}

\begin{figure}[!t]
\centering
\includegraphics[width=1\linewidth]{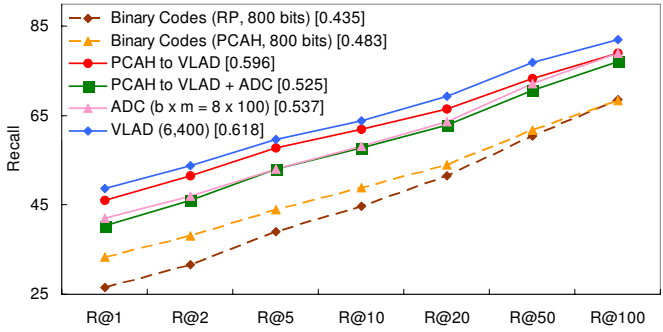}
\caption{Performance comparison of the original retrieved features (e.g., binary codes) and the approximated features on Stanford mobile visual search dataset. The approximated VLAD (PCAH to VLAD) can achieve competitive retrieval accuracy as the original VLAD. Moreover, we can combine the approximated VLAD with asymmetric distance computation (ADC) \cite{Jegou11PQ} to further reduce the memory consumption on the server-side. \textcolor{black}{The values in the square brackets represent NDCG scores.}} \label{exp:SMVS_binary2vlad}
\end{figure}

\subsection{VLAD Approximation from Reversible Binary Codes} \label{sec:exp_binary}

As demonstrated in the previous section, BoW reconstruction from binary codes might not be able to provide a good reconstruction results. Therefore, we investigate the intermediate step---VLAD approximation---in this section. As Figure~\ref{exp:SMVS_binary2vlad} shows, the approximated VLAD from binary codes (PCAH to VLAD \textcolor{black}{[0.596]}) has competitive performance as the original VLAD \textcolor{black}{[0.618]} and is better than binary codes (PCAH, 800 bits \textcolor{black}{[0.483]}). The results confirm again that we should utilize the computing power of remote servers to generate a better feature representation for image retrieval rather than using the original received feature. Another alternative is to increase the number of bits for binary codes with random projection (RP). Based on our experiments, when we utilize 6,400 bits with RP, the performance is similar to the approximated VLAD. However, it might be hard to obtain approximated VLAD from RP so we cannot further perform BoW reconstruction. By using PCA-based hashing methods, we can have better reconstruction results when increasing the number of bits and we will demonstrate the results in next section.

We also compare and integrate our proposed method with asymmetric distance computation (ADC) as adopted in the original VLAD paper \cite{Jegou12}. They compress VLAD into binary codes for database images and utilize the original VLAD for the query image to calculate the distance. Hence, they can greatly reduce the memory consumption in the database while retaining similar performance as the original approach. As Figure \ref{exp:SMVS_binary2vlad} shows, the performance of ADC method with 256 (b=8, $2^8$) centers for 100 sub-vectors (m=100, $VLAD_i$) only slightly decreases. However, by using ADC method, we still need to transmit the original VLAD to the remote server. Conversely, the proposed reversible binary codes (PCAH to VLAD) can achieve better retrieval accuracy. For fair comparison of memory usage on the server-side, we adopt the same approach as ADC that compresses database images into binary codes (8x100 bits) but we utilize the approximated VLAD as query (PCAH to VLAD + ADC). It shows that we can achieve similar recall rate as the original ADC while only transmitting few bits \textcolor{black}{(0.525 vs. 0.537)}. Moreover, the ADC approach can further combine with inverted indexing as proposed in~\cite{Jegou11PQ}.

\begin{table}[!t]
\caption{Comparison of retrieval accuracy with different bit lengths and contextual cues. The MAP of VLAD and BoW is 0.371 and 0.483, respectively. It shows that joint binary hashing (Joint PCAH-RR) can perform better when the reduced dimension is low. Moreover, by utilizing contextual information for BoW reconstruction, we can achieve similar retrieval accuracy as the original BoW while only transmitting few bits and consuming a small amount of memory on mobile devices. See more explanations in Section~\ref{sec:exp_bow}. `G' represents we utilize GPS information in the reconstruction step whereas 'B' only considers the original binary ranking results. `Bin.' stands for the final ranking results are conducted from binary codes and others are ranked by BoW.} \label{table:numberOfBits}
\vspace{-10pt}
\begin{center}
\begin{tabular}[c]{l||c|c|c|c|c}
\hline
Oxford Dataset (Bits) & 1,000 & 2,000 & 4,000 & 8,000 & 12,800  \\
\hline \hline %
Joint PCAH-RR [Bin.]      & \textbf{0.323} & 0.335 & 0.344 & -   & -  \\ 
Ind. PCAH [Bin.]   & 0.302 & \textbf{0.354} & \textbf{0.368} & \textbf{0.382} & 0.370 \\ %
Shared PCAH [Bin.]        & 0.252 & 0.306 & 0.346 & 0.375 & \textbf{0.390} \\ \hline \hline 
Joint to BoW & 0.291 & 0.314 & 0.314 & - & - \\ 
Ind. to BoW   & 0.121 & 0.200 & 0.267 & 0.312 & 0.336 \\ 
Shared to BoW & 0.066 & 0.138 & 0.233 & 0.313 & 0.343 \\ \hline 
Joint to BoW [G] & \textbf{0.398} & \textbf{0.410} & \textbf{0.413} & - & - \\ 
Ind. to BoW [G] & 0.267 & 0.314 & 0.364 & 0.390 & 0.405 \\ 
Shared to BoW [G] & 0.231 & 0.276 & 0.341 & 0.403 & 0.437 \\  \hline 
Joint to BoW [B] & 0.362 & 0.378 & 0.384 & - & - \\  
Ind. to BoW [B] & 0.240 & 0.307 & 0.363 & 0.404 & 0.404 \\ 
Shared to BoW [B] & 0.194 & 0.257 & 0.337 & 0.382 & 0.421 \\ \hline 
Joint to BoW [G+B] & 0.373 & 0.397 & 0.406 & - & - \\  
Ind. to BoW [G+B] & 0.228 & 0.309 & 0.380 &\textbf{0.419} & 0.431 \\ 
Shared to BoW [G+B] & 0.184 & 0.249 & 0.350 & 0.394 & \textbf{0.439} \\  \hline \hline 
Joint [G+B] + BRPK [B] & \textbf{0.423} & 0.456 & 0.445 & - & - \\ 
Ind. [G+B] + BRPK [B] & 0.400 & \textbf{0.465} & \textbf{0.469} & \textbf{0.481} & 0.454 \\ 
Shared [G+B] + BRPK [B] & 0.385 & 0.430 & 0.442 & 0.468 & \textbf{0.455} \\ \hline 
\end{tabular}
\end{center}
\end{table}

\subsection{BoW Reconstruction on Oxford Buildings Dataset} \label{sec:exp_bow}

Besides conducting experiments on SMVS dataset, we also evaluate our proposed method on Oxford buildings dataset. We not only compare the results by using different bit lengths and binarization methods but also utilize different contextual information (e.g., binary ranking information, GPS) for context-aware BoW reconstruction. As shown in the second to the fourth rows of Table \ref{table:numberOfBits}, the traditional binary hashing (Joint PCAH-RR [Bin.]) can perform well in low dimension (e.g., 1,000); however, the reduced dimension is limited to the number of images or features.\footnote{Although VLAD has 12,800 dimensions, the total number of Oxford dataset only contains 5,062 images. Therefore, we cannot generate more than 5,061 dimensions by standard PCA hashing unless we apply random Fourier feature (RFF) mapping \cite{Rahimi07} as adopted in \cite{Gong12}.} This phenomenon is also demonstrated in \cite{Jegou12PCA,Jegou11PQ} that joint dimension reduction will provide better compact representations and retrieval performance. As the number of bits increases (1,000 to 12,800 bits), independent binary hashing (Ind. PCAH [Bin.]) and shared binary hashing (Shared PCAH [Bin.]) can have competitive or even better retrieval accuracy than the traditional method. Note that the MAP might be low on Oxford; nevertheless, it is still similar to \cite{Jegou12} (i.e., VLAD with 64 VWs: 0.304, PCA with 128-d: 0.257).

\begin{table}[!t]
\textcolor{black}{
\caption{Comparison of the memory cost on mobile devices and retrieval accuracy on Oxford dataset. `*' indicates each dimension only consumes 8 bits.} \label{table:comparison}
\begin{center}
\begin{tabular}[c]{l||c|c|c}
\hline%
Bytes         & Transmission & Memory & MAP \\ \hline \hline %
BoW (1M) \cite{Chen09}  & $\sim$5.4K & 136M* & - \\
BoW (1M) [uncompressed] & 13.4K      & 569M  & 0.483 \\ \hline
VLAD (12,800)           & 51.2K      &  56K  & 0.371 \\
Binarized VLAD \cite{Perronnin10}  & 1.6K      &  56K  & 0.331 \\
REVV \cite{Chen11b}    & - (0.35K)  & 264K* & - \\
Ind. PCAH [Bin.] (2,000 bits)        & 0.25K      & 1080K & 0.354 \\  \hline 
Shared PCAH [Bin.] (2,000 bits)    &  0.25K      & 66K  & 0.306 \\
Shared PCAH [Bin.] (12,800 bits)   &  1.6K      & 122K  & 0.390 \\ 
Shared to BoW [B]     &  1.6K      & 122K  & 0.421 \\ \hline
\end{tabular}
\end{center}
}
\end{table}



As reported in the prior section, the approximated VLAD (PCAH to VLAD) can achieve similar results as the original VLAD. Hence, we only focus on BoW reconstruction (via VLAD) from binary codes on Oxford dataset. As shown in the fifth to the seventh rows of Table \ref{table:numberOfBits}, for fair comparison, we show the results without utilizing contextual information. The reconstructed BoW from joint PCAH-RR (Joint to BoW) is better than others because each bit is generated from (and can represent) a high-dimensional projection vector (i.e., 12,800-d) whereas other methods only consider (reconstruct) few dimensions (i.e., 128-d). However, it might consume too much memory usage on mobile devices. An alternative way is to utilize independent or shared binary hashing and increase the number of bits (still few bits for transmission). As the number of bits is 8,000 or 12,800, we can achieve similar results as the joint one.

To mitigate the loss in binarization and achieve better results, we further utilize different contextual information (i.e., [G]PS, [B]inary ranking results) for context-aware BoW reconstruction. As Table \ref{table:numberOfBits} shows, the BoW reconstruction results are much better (i.e., better than the original VLAD: 0.371), and only slightly below the original BoW (0.483). This means that we only transmit few bits and consume a small amount of memory on mobile devices to achieve competitive results as BoW (especially for challenging object queries). Moreover, as mentioned in Section \ref{subsec:reconstruction_prior}, we can also utilize pseudo-BoW from top-ranked results to roughly estimate possible VWs for a given query (binary codes). As shown in the last three rows of Table \ref{table:numberOfBits}, we can further improve the retrieval accuracy by utilizing prior knowledge from the initial binary ranking results (BRPK [B]). However, if the top-ranked images are irrelevant to the target query, as the binary ranking results in Figure \ref{fig:main_idea}, the improvement by utilizing this method might be limited.

We find that shared PCAH might be the most suitable way for MVS because it only consumes 121.9KBytes (projection matrix: 128-d x 128 bits x 4 bytes + hierarchical tree: 128-d x (10+100) x 4 bytes) on mobile devices and 1.6KBytes (12,800 bits) for transmission cost. \textcolor{black}{As Table~\ref{table:comparison} shows, for fair comparison, we only utilize image content for retrieval and compare memory cost in an uncompressed manner. It shows that BoW method can achieve the best retrieval accuracy; however, it consumes more memory and transmission cost. VLAD-like approaches can greatly reduce memory consumption while the accuracy may slightly decrease. Note that Ind. PCAH can be viewed as a simplified version of REVV.} Hence, we only increase a small amount of memory and transmission cost for better reconstruction results and retrieval accuracy.

\begin{table}
\textcolor{black}{
\caption{The retrieval accuracy on Oxford dataset with different training data. As demonstrated in prior work, the performance will slightly decrease when we train the vocabulary on different datasets.} \label{table:MAP_paris}
\vspace{-10pt}
\begin{center}
\begin{tabular}{l||c|c|c}
\hline
 & \multicolumn{3}{c}{Training data} \\
MAP & Oxford & Paris & Landmarks \\ \hline
BoW (1M) & 0.483 & 0.411 & 0.404 \\ 
VLAD (12,800) & 0.371 & 0.378 & 0.354 \\ 
Binarized VLAD \cite{Perronnin10} (100 x 129 bits) & 0.331 & 0.338 & 0.314 \\ 
Shared PCAH [Bin.] (12,800 bits) & 0.390 & 0.374 & 0.353 \\
Shared to BoW [G+B] + BRPK [B] & 0.455 & 0.391 & 0.373 \\ \hline
\end{tabular}
\end{center}
}
\end{table}

\textcolor{black}{We further utilize two additional datasets---Paris \cite{Philbin08} and Landmarks \cite{Babenko14}---for evaluating the effect of training data. As shown in Table~\ref{table:MAP_paris}, the retrieval accuracy slightly decreases when the vocabulary is trained on other datasets. This phenomenon has been demonstrated in prior work. Besides, similar to \cite{Perronnin10}, the results of binarized VLAD are slightly below the original VLAD. In our work, we assume that we can roughly \textcolor{black}{replace the original local features with BoW centers for VLAD generation.} Hence, we choose the best vocabulary tree (better BoW centers), and evaluate the effect of PCA hashing (projection matrix in Section~\ref{sec:binary2VLAD}) and reconstruction on independent datasets. As shown in Table~\ref{table:MAP_pca}, we can achieve similar retrieval accuracy as training on the database images (Oxford). Based on these experiments, we observe that a suitable vocabulary tree is essential. Hence, we can further apply the concept of fine quantization \cite{Mikulik13} or vocabulary adaptation \cite{arandjelovic13} to our proposed method.}

\textcolor{black}{For retrieval time, as demonstrated in prior work, binary matching and BoW matching with inverted indexing are very efficiency for real-time retrieval systems. Hence, in our experiments, we consume more time on the reconstruction step. From binary codes to approximated VLAD, it only contains a matrix multiplication. However, the BoW reconstruction step with sparse constraint (Eq.~\eqref{eq:formulation_BoW_sparsity}) takes around 0.14 seconds for each $VLAD_i$ (128-d) when we consider all the possible VWs. To further apply dictionary selection, we can reduce it to 0.05s (i.e., around 5s per query). For integrating with prior knowledge (Section~\ref{subsec:reconstruction_prior}), we relax the sparse constraint and thus reduce the reconstruction time to 0.51s per query. Moreover, take advantages of cloud, we can parallel the reconstruction step for each $VLAD_i$ and utilize multiple cloud servers to further reduce the reconstruction time.}

\begin{table}
\caption{The retrieval accuracy on Oxford dataset with different training data on PCA hashing. We can achieve similar performance as training on Oxford} \label{table:MAP_pca}
\begin{center}
\begin{tabular}{l||c|c|c}
\hline
 & \multicolumn{3}{c}{Training data} \\
MAP & Oxford & Paris & Landmarks \\ \hline
VLAD (12,800) & \multicolumn{3}{c}{0.371} \\ \hline 
Shared PCAH [Bin.] (12,800 bits) & 0.390 & 0.393 & 0.384 \\
Shared to BoW [G+B] (12,800 bits) & 0.439 & 0.430 & 0.433 \\
Shared to BoW [G+B] + BRPK [B] & 0.455 & 0.450 & 0.449 \\ \hline
BoW (1M) & \multicolumn{3}{c}{0.483} \\ \hline 
\end{tabular}
\end{center}
\end{table}

\section{Conclusions and future works}

In this work, we propose context-aware BoW reconstruction that utilizes the computing power of remote servers for mobile visual search. We focus on generating reversible and memory-efficient binary codes on mobile devices, and attempt to reconstruct them to a better BoW representation on remote servers (cloud). Hence, the proposed method only transmits few bits to the remote server. By observing the relation between VLAD and BoW, we can reconstruct BoW from VLAD or binary codes. Moreover, we can select possible visual features (VWs) according to the contextual information (e.g., top-ranked images, category, GPS), and further incorporate with prior knowledge from the initial (binary) ranking results. Experimental results show that the proposed method can achieve better retrieval results compared to the original retrieved feature (e.g., VLAD or binary codes). In the future, we will investigate how to utilize extra information on the server-side and adopt better binary reconstruction methods such as \cite{Hinton06}.

\ifCLASSOPTIONcaptionsoff
  \newpage
\fi



\bibliographystyle{IEEEtran}
\bibliography{references}
\end{document}